\documentclass[aps, prl, notitlepage, citeautoscript, superscriptaddress, reprint]{revtex4-2}

\usepackage{xr}
\usepackage{amsmath}
\usepackage{amsfonts, amssymb,amsxtra}
\usepackage[]{graphicx}
\usepackage{grffile}
\usepackage{amsfonts}
\usepackage{framed}
\usepackage{bbm}
\usepackage{braket}
\usepackage{xcolor}
\usepackage{mathtools}

\usepackage[normalem]{ulem}

\usepackage{pdfpages}
\makeatletter
\AtBeginDocument{\let\LS@rot\@undefined}
\makeatother

\definecolor{forestgreen}{rgb}{0.13, 0.55, 0.13}

\usepackage[colorlinks=true]{hyperref}
\hypersetup{
    unicode=false,          % non-Latin characters
    pdftoolbar=true,        % show Acrobat
    pdfmenubar=true,        % show Acrobat
    pdffitwindow=false,     % window fit to page when opened
    pdfstartview={FitH},    % fits the width of the page to the window
    pdftitle={},    % title
    pdfauthor={},     % author
    pdfsubject={},   % subject of the document
    pdfcreator={},   % creator of the document
    pdfproducer={}, % producer of the document
    pdfkeywords={} {} {}, % list of keywords
    pdfnewwindow=true,      % links in new window
    colorlinks=true,       % false: boxed links; true: colored links
    linkcolor=magenta, %red,          % color of internal links (change box color with linkbordercolor)
    citecolor=blue,        % color of links to bibliography
    filecolor=magenta,      % color of file links
    urlcolor=blue           % color of external links
}

\begin{document}
\title{The essential role of magnetic frustration in the phase diagrams of doped cobaltites}

\author{Peter P.~Orth}
\email{porth@iastate.edu}
\affiliation{Ames Laboratory, Ames, Iowa 50011, USA}
\affiliation{Department of Physics and Astronomy, Iowa State University, Ames, Iowa 50011, USA}

\author{Daniel Phelan}
\affiliation{Materials Science Division, Argonne National Laboratory, Argonne, Illinois 60439, USA}
\affiliation{Department of Chemical Engineering and Materials Science, University of Minnesota, Minneapolis, MN 55455, USA}

\author{J. Zhao}
\affiliation{Materials Science Division, Argonne National Laboratory, Argonne, Illinois 60439, USA}

\author{H. Zheng}
\affiliation{Materials Science Division, Argonne National Laboratory, Argonne, Illinois 60439, USA}

\author{J. F. Mitchell}
\affiliation{Materials Science Division, Argonne National Laboratory, Argonne, Illinois 60439, USA}

\author{C. Leighton}
\affiliation{Department of Chemical Engineering and Materials Science, University of Minnesota, Minneapolis, MN 55455, USA}

\author{Rafael M. Fernandes}
\affiliation{School of Physics and Astronomy, University of Minnesota, Minneapolis,
Minnesota 55455, USA}

\begin{abstract}
   Doped perovskite cobaltites (e.g., La$_{1-x}$Sr$_x$CoO$_3$) have been extensively studied for their spin-state physics, electronic inhomogeneity, and insulator-metal transitions. Ferromagnetically-interacting spin-state polarons emerge at low $x$ in the phase diagram of these compounds, eventually yielding long-range ferromagnetism. The onset of long-range ferromagnetism ($x \approx 0.18$) is substantially delayed relative to polaron percolation ($x \approx 0.05$), however, generating a troubling inconsistency.
   Here, Monte-Carlo simulations of a disordered classical spin model are used to establish that previously ignored \emph{magnetic frustration} is responsible for this effect, enabling faithful reproduction of the magnetic phase diagram.
\end{abstract}

\date{\today}
\maketitle

\emph{Introduction and experimental situation.--}
The insulator-metal transition (IMT) is an important, widely observed phenomenon in condensed matter physics that continues to be intensively studied~\cite{imadaMetalinsulatorTransitions1998a}. While there are various mechanisms by which IMTs occur, in quantum materials, percolation of conductive regions in an insulating matrix, driven by temperature, doping, pressure, \emph{etc.}, is widespread~\cite{imadaMetalinsulatorTransitions1998a,dassarmaTwoDimensionalMetalInsulatorTransition2005, meirPercolationTypeDescriptionMetalInsulator1999,zhangDirectObservationPercolation2002,wardElasticallyDrivenAnisotropic2009}. The microscopic factors inducing such electronic inhomogeneity (across nano- to meso-scales) include structural disorder, multiple competing interactions, and coupled degrees of freedom, resulting in a formidable problem~\cite{imadaMetalinsulatorTransitions1998a,dagottoComplexityStronglyCorrelated2005,tokuraCorrelatedElectronPhysicsTransitionMetal2003}.

Doped perovskite cobaltites, with La$_{1-x}$Sr$_x$CoO$_3$ (LSCO) as the archetype, have proven to be paradigmatic for investigation of such percolative IMTs~\cite{senaris-rodriguezMagneticTransportProperties1995,wuGlassyFerromagnetismMagnetic2003,heDopingFluctuationdrivenMagnetoelectronic2009,wuIntergranularGiantMagnetoresistance2005,phelanNanomagneticDropletsImplications2006}. Experimental studies of LSCO single crystals culminate in the electronic/magnetic phase diagram in Fig.~\ref{fig:figure_1}, which we have constructed from both published and new data (see Supplemental Material, Sec.~S.I~\footnote{The Supplemental Material contains details on the construction of the phase diagram in Fig.~1 (Sec.~S.I), on the doping evolution of the low temperature specific heat (Sec.~S.II)., and on the results of the F Binder cumulant in the theoretical model (Sec.~S.III).}). At $x = 0$ [undoped LaCoO$_3$ (LCO)], the Co$^{3+}$ ($d^6$) ions adopt an $S = 0$ diamagnetic insulating ground state, but with a spin gap of only $\sim 10$~meV due to comparable crystal field and Hund’s exchange energies, leading to the famous thermally-excited spin-state transition (SST)~\cite{wuGlassyFerromagnetismMagnetic2003, heDopingFluctuationdrivenMagnetoelectronic2009,wuIntergranularGiantMagnetoresistance2005,phelanNanomagneticDropletsImplications2006,podlesnyakSpinStateTransitionMathrmLaCoO2006,haverkortSpinStateTransition2006,korotinIntermediatespinStateProperties1996,noguchiEvidenceExcitedTriplet2002,raccahFirstOrderLocalizedElectronEnsuremath1967}. Defined as the mid-point of the resulting rise in susceptibility, the SST occurs at $T_{\text{SST}} \approx 70$~K (Fig.~\ref{fig:figure_1}). It is essentially complete by approximately $120$~K, above which antiferromagnetic (AF) Co$^{3+}$-Co$^{3+}$ superexchange interactions occur, resulting in the negative Curie-Weiss temperature ($\theta_{\text{CW}}$)~\cite{jonkerMagneticCompoundsWtth1953,jonkerMagneticSemiconductingProperties1966,podlesnyakEffectCarrierDoping2011} shown in Fig.~\ref{fig:figure_1}.

\begin{figure}[h!]
    \centering
    \includegraphics[width=.9\linewidth]{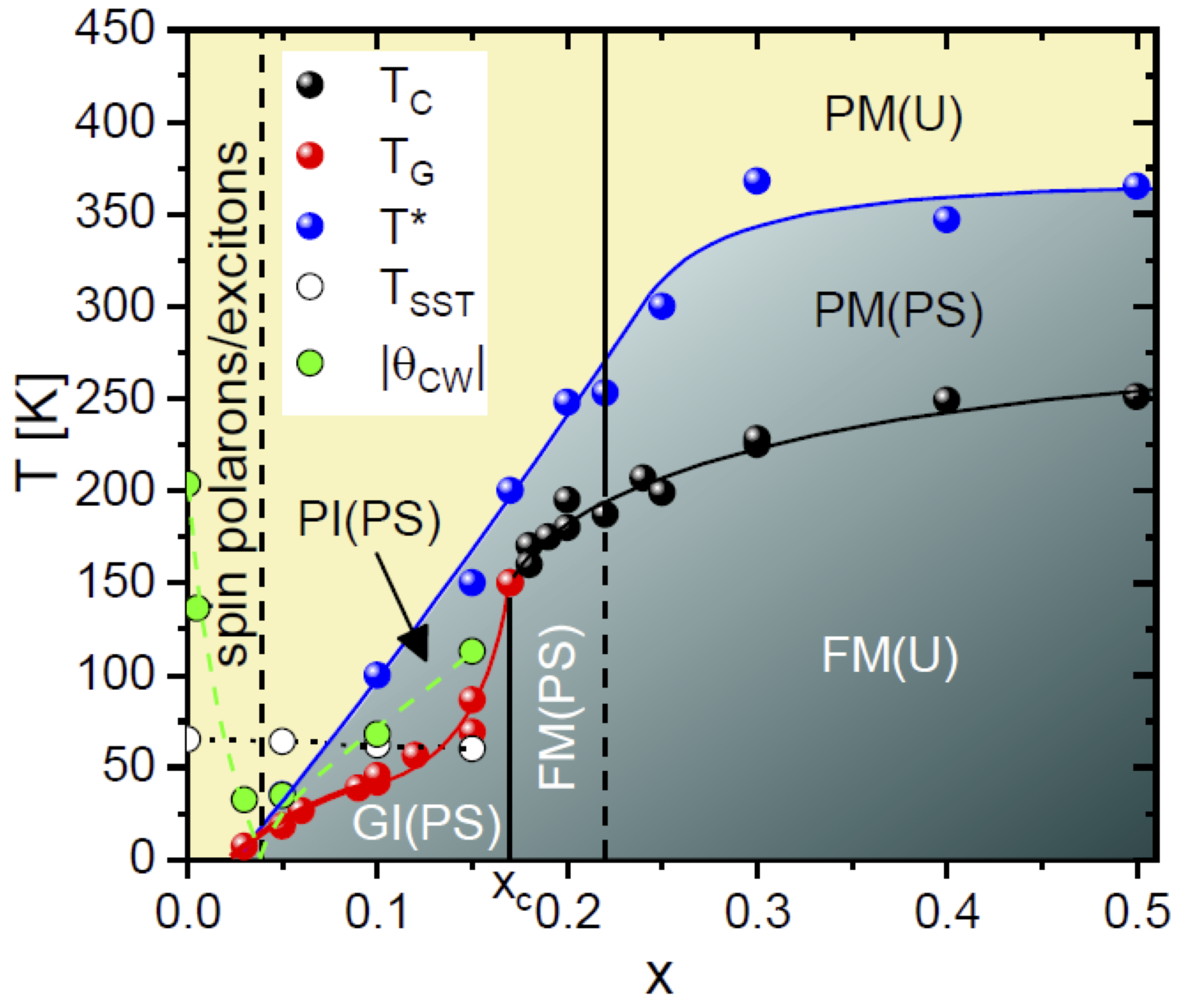}
    \caption{
    Magnetic/electronic phase diagram of single crystal La$_{1-x}$Sr$_x$CoO$_3$ (LSCO). Shown are the Curie temperature ($T_{\text{C}}$), spin/cluster glass freezing temperature ($T_{\text{G}}$), onset temperature for ferromagnetic (F) fluctuations/clusters ($T^*$), and non-F matrix spin-state crossover temperature ($T_{\text{SST}}$).
    Superimposed is the magnitude of the Curie-Weiss temperature $\theta_{\text{CW}}$, which inverts from antiferromagnetic (AF) to F at $x \approx 0.04$. FM = ferromagnetic metal, PM = paramagnetic metal, PI = paramagnetic insulator, and GI = glassy magnetic insulator; U designates uniform states, PS magnetically/electronically phase separated states.
    $T_{\text{C}}$, $T_{\text{G}}$, $T^*$ and $\theta_{\text{CW}}$ are from this work, supplemented with Refs.~\cite{heDopingFluctuationdrivenMagnetoelectronic2009,heHeatCapacityStudy2009,heMagnetoelectronicPhaseSeparation2009,smithSpinPolaronsTextLa2008}; $T_{\text{SST}}$ data are taken from Ref.~\cite{smithEvolutionSpinstateTransition2012}. The vertical line at $x_c = 0.18$ marks the cluster percolation threshold, the vertical line at $x = 0.22$ the transition from PS to U states at low temperature (and from negative to positive temperature coefficient of resistance at high temperature). For $x > 0.30$, where single crystal data are not available, $T_\text{C}$ and $T^*$ values from polycrystalline samples have been used~\cite{heNonGriffithslikeClusteredPhase2007}.
    }
    \label{fig:figure_1}
\end{figure}

\begin{figure}[t!]
    \centering
    \includegraphics[width=\linewidth]{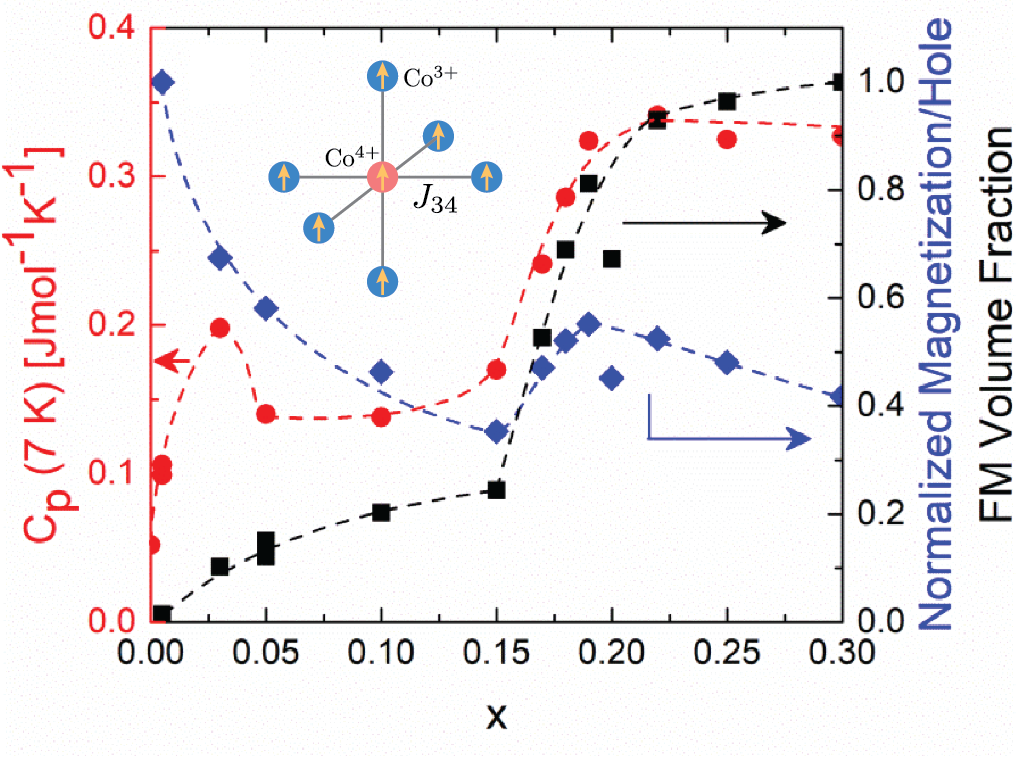}
    \caption{Doping ($x$) dependence of the 7 K heat capacity ($C_p$, red, left axis), magnetization per hole (blue, right axis, normalized to the $x = 0$ value), and F volume fraction (black, right axis). The latter is estimated by extrapolating the high field magnetization-field behavior to zero field. Dashed lines are guides to the eye. Inset: Schematic of seven-site spin-state polaron with F exchange coupling $J_{34}$ between sites.}
    \label{fig:figure_2}
\end{figure}

Fascinating behavior emerges upon dilute hole doping (e.g., $x = 0.005$), where giant magnetic moments ($S \approx 13/2$) occur due to seven-site, octahedrally-shaped spin-state polarons (Fig.~\ref{fig:figure_2}, inset)~\cite{yamaguchiSpinstateTransitionHighspin1996,podlesnyakSpinStatePolaronsLightlyHoleDoped2008,podlesnyakEffectCarrierDoping2011,alfonsovOriginSpinstatePolaron2009}. In essence, doped Co$^{4+}$ ions stabilize finite spin states on neighboring Co$^{3+}$ ions, forming a spin-state polaron with ferromagnetic (F) intra-polaron interactions.
As noted previously, and borne out by our own data (Fig.~\ref{fig:figure_2}, right axis), increasing $x$ leads to polaron overlap~\cite{yamaguchiSpinstateTransitionHighspin1996}, and thus collapse of the magnetization per hole~\cite{podlesnyakEffectCarrierDoping2011}. As shown in Fig.~\ref{fig:figure_1}, this is accompanied by a rapid increase in $\theta_{\text{CW}}$ (decrease in $|\theta_{\text{CW}}|$), which inverts from negative to positive~\cite{jonkerMagneticCompoundsWtth1953,podlesnyakEffectCarrierDoping2011} at $x \approx 0.04$, reflecting dominance of intra-polaron F Co$^{4+}$-Co$^{3+}$ interactions over extra-polaron AF Co$^{3+}$-Co$^{3+}$ interactions. At this pivotal $x \approx 0.04$ (vertical dashed line, Fig.~\ref{fig:figure_1}) multiple experimental probes indicate short-range F order, including small-angle neutron scattering (SANS)~\cite{heDopingFluctuationdrivenMagnetoelectronic2009}, inelastic neutron spectroscopy~\cite{podlesnyakEffectCarrierDoping2011}, neutron diffraction~\cite{phelanNanomagneticDropletsImplications2006,phelanSpinIncommensurabilityTwo2006}, and specific heat~\cite{heHeatCapacityStudy2009,heDopingFluctuationdrivenMagnetoelectronic2009}. As shown in the phase diagram, signatures of nanoscale magnetic inhomogeneity then turn on at temperature $T^*$, followed at a lower $T_G$ by spin/cluster-glass freezing. Most dramatically, a percolation transition then occurs at $x_c \approx 0.18$ (solid vertical line, Fig.~\ref{fig:figure_1}), where a low temperature IMT occurs and long-range F order is detected by neutron diffraction~\cite{satheNeutronDiffractionStudies1996,phelanNanomagneticDropletsImplications2006}, SANS~\cite{wuIntergranularGiantMagnetoresistance2005,heDopingFluctuationdrivenMagnetoelectronic2009}, and magnetometry~\cite{wuGlassyFerromagnetismMagnetic2003,senaris-rodriguezMagneticTransportProperties1995}. In Fig.~\ref{fig:figure_2} this is reflected in a sharp upturn in magnetization (blue, right axis) and F volume fraction (black, right axis). This percolation transition can also be controlled with voltage in electrolyte-gated LSCO~\cite{orthPercolationCombinedElectrostatic2017,walterGiantElectrostaticModification2018}. A final important composition on the phase diagram is at $x \approx 0.22$ (vertical line, Fig.~\ref{fig:figure_1}), where multiple probes (e.g., SANS, specific heat, La NMR, magnetotransport~\cite{heDopingFluctuationdrivenMagnetoelectronic2009,smithSpinPolaronsTextLa2008}) reveal \emph{uniform} long-range F order, i.e., an end to the low $T$ magnetically-phase-separated regime~\cite{heDopingFluctuationdrivenMagnetoelectronic2009}. $T_C$ gradually increases with further doping, reaching $250$~K at $x = 0.5$.

The picture that emerges from the above, \emph{i.e.}, spin-state polarons generating nanoscale F regions that percolate into long-range F order at $x_c \approx 0.18$, has become widely accepted. This masks a troubling inconsistency, however. Specifically, simple statistical arguments indicate that percolation of seven-site polarons into a long-range F network should occur at $x \approx 0.04$. This can be understood by dividing the cubic (LSCO is cubic above $x \approx 0.5$ and mildly rhombohedrally distorted at lower $x$) site percolation threshold of $s_c = 0.31$ by 7 (the number of sites in the polaron), yielding $x = s_c/7 = 0.044$. The number of \emph{isolated} polarons in fact peaks at $x \approx 0.03$ (see Fig.~\ref{fig:figure_3}(b), inset), prior to percolation of the polarons at $x \approx 0.04$. As discussed in the Supplemental Material, Sec.~S.II~\cite{Note1}, the low $T$ specific heat $C_p$ (e.g., at $7$~K, left axis, Fig.~\ref{fig:figure_2}) supports this, the Schottky specific heat signature of the spin-state polarons peaking at $x = 0.03$ then leveling off at $x = 0.05$, before rising again at $x > x_c$ due to electronic contributions to $C_p$ in the F metallic state~\cite{heDopingFluctuationdrivenMagnetoelectronic2009,heHeatCapacityStudy2009}.
Critically, this doping level at which polaron percolation is expected ($x \approx 0.04$) is substantially lower than the experimentally observed $x_c \approx 0.18$ at which percolation to a long-range F metallic state occurs. In this work, we finally resolve this issue, using a model of interacting dilute and disordered magnetic moments that we investigate using large-scale Monte-Carlo simulations. This reveals a vital role for \emph{magnetic frustration}, a previously ignored factor in the phase diagrams of such materials. We expect this physics to also play a role in the phase behavior of other heavily-studied doped cobaltites, such as La$_{1-x}$Ca$_x$CoO$_3$, La$_{1-x}$Ba$_x$CoO$_3$, Pr$_{1-x}$Ca$_x$CoO$_3$, etc. Moreover, while the spin-state aspect to the polarons is specific to cobaltites, frustration may also play a role in other systems exhibiting magnetic polaron to long-range order transitions, such as manganites.

\emph{Theoretical model.--}
To address the question of what delays the onset of percolation in LSCO, we study an effective disordered classical spin model that captures the essential microscopic physics.
\begin{figure}[tb]
    \centering
    \includegraphics[width=\linewidth]{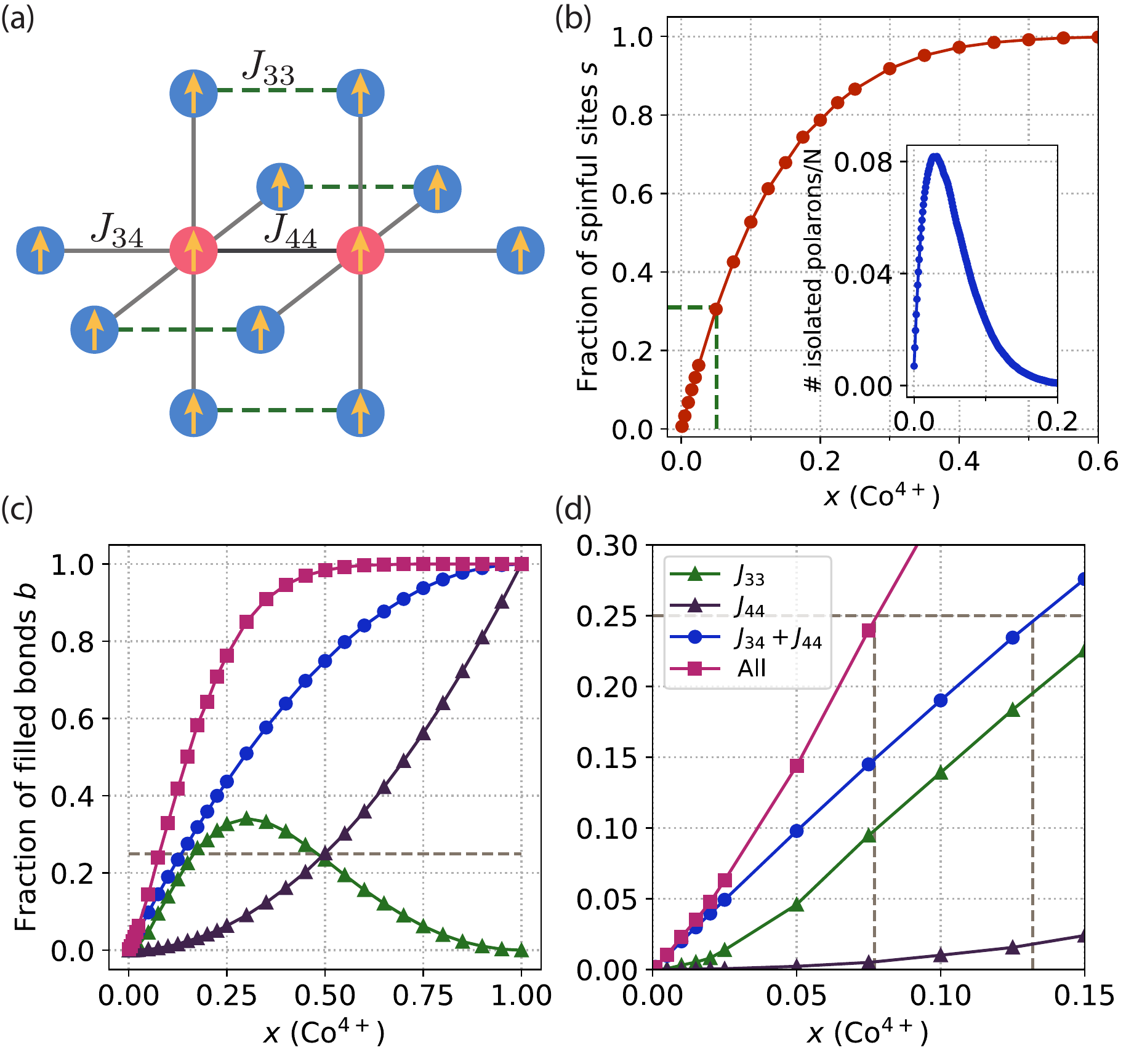}
    \caption{(a) Schematic of two nearby $7$-site polarons with exchange interactions $J_{34}, J_{44} < 0$ (F) and $J_{33} > 0$ (AF). Co$^{4+}$ (Co$^{3+}$) sites are shown in red (blue). In this F configuration, the AF $J_{33}$ bonds (green dashed) are frustrated.
    (b) Fraction of spinful sites $s$ (either Co$^{4+}$ or Co$^{3+}$) as a function of Co$^{4+}$ doping $x$. The horizontal green dashed line denotes the site percolation threshold on the cubic lattice, $s_c = 0.31$, which is reached at $x = 0.05$. Inset shows the number of isolated $7$-site polarons, which peaks at $x=0.03$.
    (c-d) Fraction of filled bonds $b$ as a function of Co$^{4+}$ doping $x$. Different bonds are described in the legend; ``All" refers to counting $J_{33}$, $J_{34}$, and $J_{44}$ bonds. The horizontal grey line denotes the bond percolation threshold on the cubic lattice $b_c = 0.25$. The system exhibits only finite size clusters for $x < 0.08$ ($b_{\text{All}} < b_c$). If AF bonds were absent, the $T=0$ paramagnetic to ferromagnetic transition would occur at $x = 0.13$ ($b_{(34)+(44)} = b_c$), when a macroscopic percolating cluster that only contains F $J_{34}$ and $J_{44}$ bonds emerges. However, the presence of AF $J_{33}$ bonds decouples the percolation of F bonds from the onset of F order.}
    \label{fig:figure_3}
\end{figure}
Starting from an empty cubic lattice, we randomly populate a fraction $x$ of the sites with Ising spins $S^{(4)}_i = \pm 1$, corresponding to Co$^{4+}$ ions. Two neighboring $S^{(4)}$ spins interact ferromagnetically with coupling constant $J_{44} < 0$, reflecting the fact that the Curie-Weiss temperature $\theta_{\text{CW}}$ is positive for $x \geq 0.05$ and the fully doped end-member SrCoO$_{3}$ is F ($T_\text{C} \approx 300$~K). Note that we consider Ising spins here to account for the easy-axis anisotropy in the material. We then place spins $S^{(3)}_i = \pm 1$ on all empty sites neighboring an $S^{(4)}$ spin, which models the emergence of spin-state polarons due to transitions from LS (low-spin) to IS (intermediate-spin) states on the Co$^{3+}$ sites neighboring a Co$^{4+}$. The nearest-neighbor interaction between $S^{(4)}_i$ and $S^{(3)}_j$ spins is also F, $J_{34} < 0$, leading to the formation of large spin $S$ polaronic clusters, consisting of 7 sites for an isolated polaron. In contrast, the interaction between two neighboring $S^{(3)}$ %(IS Co$^{3+}$)
spins is AF, $J_{33} > 0$. This is consistent with a negative $\theta_{\text{CW}}$ for thermally excited IS Co$^{3+}$ spins at $x < 0.05$.

The presence of both F and AF interactions leads to frustration of polarons as shown in Fig.~\ref{fig:figure_3}(a) for two neighboring polarons. In the F configuration shown, the F bonds $J_{34}$ and $J_{44}$ are satisfied, thus frustrating the AF bonds $J_{34}$. The fact that the number of spinful sites $s$ rapidly grows with the fraction $x$ of Co$^{4+}$ sites is shown in Fig.~\ref{fig:figure_3}(b). Since an isolated polaron contains $7$ sites, the initial slope of the curve is $s \approx 7 x$, which levels off once the polarons overlap at larger $x$. The maximal number of isolated polarons occurs at $x = 0.03$ (Fig.~\ref{fig:figure_3}(b), inset). This marks the onset of frustration, which suppresses the development of F order once a macroscopic site cluster forms at $x = 0.05$, where $s$ reaches the site percolation threshold $s_c = 0.31$. Note that we neglect the weak correlation between the occupation probability of spinful sites as they are added as few-site polaronic clusters, which will result in a slightly larger numerical value of $s_c$. To further elucidate frustration, we investigate the phenomenological model Hamiltonian
\begin{align}
H &= J_{44} \sum_{\langle i, j \rangle_{44}} S^{(4)}_i S^{(4)}_j  + J_{33} \sum_{\langle i, j \rangle_{33}} S^{(3)}_i S^{(3)}_j \nonumber \\
& \qquad + J_{34} \sum_{\langle i, j \rangle_{34}} S^{(3)}_i S^{(4)}_j  \,.
\label{eq:Hamiltonian}
\end{align}
Here, $J_{44}, J_{34} < 0$ correspond to F and $J_{33} > 0$ to AF interactions and $\langle i, j \rangle_{ab}$ runs over all nearest-neighbor spinful sites of type $a,b$ on the cubic lattice.

\begin{figure}[h!]
    \centering
    \includegraphics[width=\linewidth]{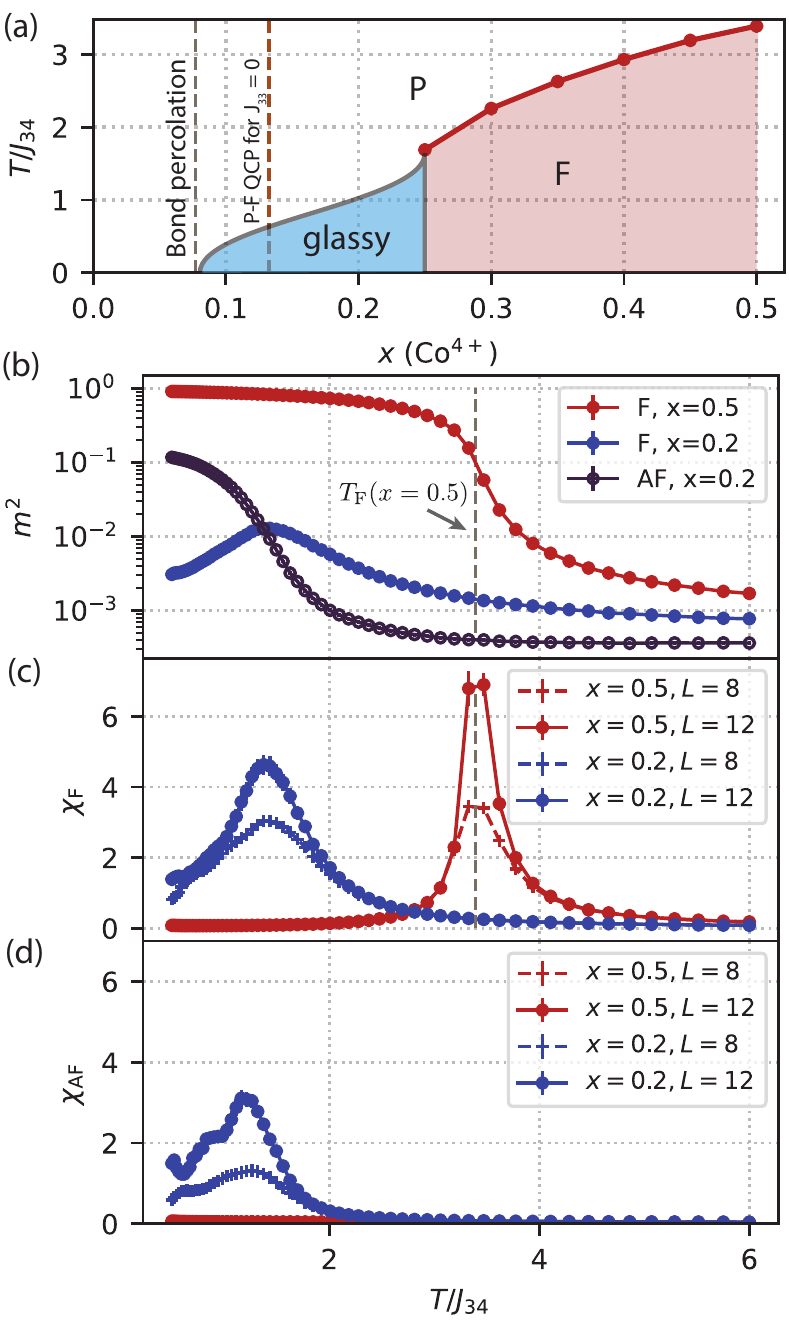}
    \caption{(a) Numerical finite-temperature phase diagram of the frustrated $J_{33}$-$J_{34}$-$J_{44}$ model as a function of Co$^{4+}$ doping, obtained from percolation analysis and MC simulations. The different phases are paramagnetic (P), ferromagnetic (F), and glassy, which is characterized by short-range AF and F correlations. Couplings are set to $J_{34}$ $=J_{44}$ $=-1$ (F) and $J_{34}=0.2$ (AF). Transition temperatures $T_\text{F}$ (red dots) are obtained from MC simulations via crossing of F Binder cumulants. Such crossings are notably absent for $x \leq 0.225$. The grey dashed line at $x = 0.08$ denotes the bond percolation threshold ($b_{\text{All}} = b_c$, see Fig.~\ref{fig:figure_3}(d)). The brown line denotes the $T=0$ P-F quantum critical point (QCP) in the unfrustrated model ($J_{33}=0$), obtained from $b_{34 + 44} = b_c$ in Fig.~\ref{fig:figure_3}(d). The transition temperature to the glassy phase (solid grey line) is schematic and smoothly connects the various numerically obtained points.
    (b-d) MC simulation results of the observables $m^2_{\text{F}}, m^2_{\text{AF}}, \chi_{\text{F}},  \chi_{\text{AF}}$ for $x=0.2$ and $x=0.5$ and different system sizes, $L=8, 12$. Long-range F order develops for $x=0.5$, while only short-range F and AF correlations coexist at $x=0.2$, which is characteristic of a glassy phase.
}
    \label{fig:figure_4}
\end{figure}

Let us first consider the limit $J_{33} = 0$, where only F interactions are present. As shown in Fig.~\ref{fig:figure_3}(c,d), counting only $J_{34}$ and $J_{44}$ bonds (blue points), the bond percolation limit at $b_c = 0.25$~\cite{staufferIntroductionPercolationTheory1994,shklovskiiElectronicPropertiesDoped1984} is reached at Co$^{4+}$ doping level $x_c^{(34) + (44)} = 0.13$ (Fig.~\ref{fig:figure_3}(d) is a magnified view of Fig.~\ref{fig:figure_3}(c) to enable this to be seen more clearly). Since all interactions in this limit are F, this implies the emergence of a macroscopic F cluster (consisting solely of F bonds) and the onset of F order at $T=0$. In contrast, considering $J_{33} \neq 0$ and counting \emph{all} bonds (magenta points), a macroscopic percolating cluster emerges already at $x_{c}^{(33) + (34) + (44)} = 0.075$. Critically, however, $J_{33} > 0$ is AF, so the development of F order does \emph{not} coincide with bond percolation. Instead, spins in the macroscopic cluster will experience frustration due to the random distribution of F and AF bonds, and we expect a spin glass (SG) phase at low temperatures. The degree of frustration is controlled by the ratios $J_{33}/J_{34}$ and $J_{33}/J_{44}$, which also set the endpoint of the SG phase and the emergence of F order at larger $x$, as we show next.

We numerically investigate the model in Eq.~\eqref{eq:Hamiltonian} using large-scale parallel tempering Monte-Carlo (MC) simulations. We simulate systems of size $N = L^3$ with $L = 8, 12, 16$ and for $N_\text{dis} = 50$ ($30$) disorder realizations for $L=8$ ($L=12, 16$) using a combination of Metropolis and parallel-tempering updating steps. The total number of MC steps is $N_{\text{MC}} = 8 \times 10^4$, $4 \times 10^4$, $8 \times 10^3$ for increasing system sizes. We use the first half of these steps for thermalization and measure observables only during the second half of the simulation. Error bars are obtained using the standard jackknife method~\cite{millerJackknifeReview1974}.
We choose the F exchange couplings to be equal $J_{34} = J_{44} = -1$, since the F transition temperature of LSCO varies only slightly from $\sim 250$ to $\sim 300$~K over the doping range $0.5 < x < 1$, whereas the bond number ratio $b_{34}/b_{44}$ varies from two to zero over the same range. Because there are only few $J_{44}$ bonds for small doping ($b_{44}<0.06$ for $x<0.25$), our results in this range depend only weakly on $J_{44}$. We set $J_{34} = 0.2$, which corresponds to moderate frustration and gives $x_c = 0.25$, close to the experimentally observed value in LSCO. A smaller $J_{34}$ merely shifts $x_c$ closer to $x= 0.13$%[see Fig.~\ref{fig:figure_3}(d)
, whereas a larger $J_{34}$ may result in an intermediate AF phase; this is in fact observed in some doped cobaltites, e.g., La$_{1-x}$Ba$_x$CoO$_3$~\cite{tongPossibleLinkStructurally2011}.

The resulting finite temperature phase diagram as a function of $x$, and for $J_{34} = J_{44} = -1$, $J_{33} = 0.2$, is shown in Fig.~\ref{fig:figure_4}(a); this is plotted over the same range of $x$ shown in the phase diagram in Fig.~\ref{fig:figure_1}.
At large $x$, where F bonds dominate, the system enters a F phase at a transition temperature $T_{\text{F}}(x)$ that is slightly smaller than the value $T^{\text{3D}}_{\text{Ising}} = 4.5\, J_{44}$ of the $x=1$ system due to missing bonds and AF $J_{33}$ bonds in the macroscopic spin cluster. We obtain $T_{\text{F}}$ from the crossing of the F Binder cumulants~\cite{binderFiniteSizeScaling1981, sandvikComputationalStudiesQuantum2010a} $U_{\text{F}} = \frac32\bigl(1 - \frac13 \frac{[\langle m^4 \rangle]_{\text{dis}}}{[\langle m^2 \rangle]^2_{\text{dis}}} \bigr)$ for different system sizes (see~\cite{Note1} for details). Here, $m = \frac{1}{N} \sum_{i=1}^N S_i$ is the magnetization, $S_i$ is the spin at site $i$, and the brackets $\langle \mathcal{O} \rangle$ ($[\mathcal{O}]_{\text{dis}}$) denote thermal (disorder) averages.

$T_{\text{F}}(x)$ decreases as $x$ is reduced and for $x \leq 0.225$ the crossing of the Binder cumulants is absent, implying that true long-range F order ceases to exist. By analogy to the Edwards-Anderson model~\cite{edwardsTheorySpinGlasses1975,katzgraberUniversalityThreedimensionalIsing2006,kawashimaRecentProgressSpin2005}, we expect the system to exhibit a magnetic-glassy multicritical point~\cite{ledoussalLocationIsingSpinGlass1988, hasenbuschCriticalBehaviorThreedimensional2007,hasenbuschCriticalBehaviorThreedimensional2008} and develop a low-T spin glass (SG) phase for $x \leq 0.225$. We qualitatively verify this scenario by investigating both F and AF observables, as shown in Fig.~\ref{fig:figure_4}(b-d).
At $x=0.5$ the system behaves like a typical ferromagnet with a peak in the F susceptibility $\chi_{\text{F}} = \frac{N}{T} \bigl( [\langle m^2 \rangle - \langle m \rangle^2 ]_{\text{dis}} \bigr)$ that grows with system size (Fig.~\ref{fig:figure_4}(c)) and a saturation magnetization $\langle m^2 \rangle$ that approaches unity as $T \rightarrow 0$ (Fig.~\ref{fig:figure_4}(b)).
In contrast, at $x = 0.2$, we find competing F and AF fluctuations with broad peaks of comparable size in $\chi_{\text{F}}$ (Fig.~\ref{fig:figure_4}(c)) and $\chi_{\text{AF}}$ (Fig.~\ref{fig:figure_4}(d)), where $\chi_{\text{AF}}$ measures the susceptibility at wavevector $q = (\pi,\pi,\pi)$. Both F and AF order parameter magnetizations are small (and finite) in such a finite system (Fig.~\ref{fig:figure_4}(b)).
We leave a detailed study of the glassy phase, and other parameter regions in this model, for future work.

We can determine the $T=0$ boundary of the glassy phase on the lower doping side from our bond percolation analysis presented above (Fig.~\ref{fig:figure_3}(c)), which gave $x_{\text{SG},\text{min}} = 0.075$. Since the SG transition temperature must vanish at this point, we obtain the phase diagram shown in Fig.~\ref{fig:figure_4}(a), which bears a striking resemblance to the experimental phase diagram in Fig.~\ref{fig:figure_1}. Note that we focus on phase transitions below the $T^*$ line, where finite order parameters develop. This is a strong indication that our model correctly captures the microscopic physics in the low and intermediate doping range in LSCO, and that it is the frustration of polarons due to competing AF and F interactions that underlies the shift of the percolation threshold from the naively expected value $x \simeq 0.05$ to the experimentally observed value $x_{c} = 0.18$.
Given the universality of the spin-state physics and spin-state polaron formation in other doped cobaltites, such as La$_{1-x}$Ca$_x$CoO$_3$, La$_{1-x}$Ba$_x$CoO$_3$, Pr$_{1-x}$Ca$_x$CoO$_3$, etc., and their similar qualitative behavior with doping, it is highly likely that this model can be applied far more broadly than to La$_{1-x}$Sr$_x$CoO$_3$.

\emph{Conclusion.--} In essence, this work reveals that a factor previously ignored in the physics of doped cobaltites - magnetic frustration - plays an inherent role in shaping their electronic/magnetic phase diagrams. Such frustration, which is of broad importance in condensed matter systems, delays the percolation associated with magnetic polarons, playing a vital role in the insulator-metal transition in these systems. This is yet another illustration of the importance of frustration in magnetic materials, in this case coupled to the ubiquitous problem of the insulator-metal transition.

\begin{acknowledgments}
We acknowledge valuable discussions with B.~I. Shklovskii and T. Vojta.
Experimental work at the University of Minnesota (UMN) was supported by the US Department of Energy through the UMN Center for Quantum Materials under DE-SC-0016371. Work at Argonne National Laboratory (crystal growth and magnetic characterization) was supported by the U.S. Department of Energy, Office of Science, Basic Energy Sciences, Materials Science and Engineering Division. During conception and execution of the theory work, R.M.F. and P.P.O. were supported by the National Science Foundation through the UMN MRSEC under Grant No. DMR-1420013. P.P.O also acknowledges support from start-up funds from Iowa State University during the late stages of the theory work.
\end{acknowledgments}

\bibliographystyle{apsrev4-2}
%\bibliography{Biblio_LSCO}


\clearpage

\pagebreak


\section*{Supplementary material}

# Supplemental Material to "The essential role of magnetic frustration in the phase diagrams of doped cobaltites"

Peter P. Orth,[1, 2, *] Daniel Phelan,[3, 4] J. Zhao,[3] H. Zheng,[3] J. F. Mitchell,[3] C. Leighton,[4] and Rafael M. Fernandes[5]

[1] *Ames Laboratory, Ames, Iowa 50011, USA*
[2] *Department of Physics and Astronomy, Iowa State University, Ames, Iowa 50011, USA*
[3] *Materials Science Division, Argonne National Laboratory, Argonne, Illinois 60439, USA*
[4] *Department of Chemical Engineering and Materials Science,
University of Minnesota, Minneapolis, MN 55455, USA*
[5] *School of Physics and Astronomy, University of Minnesota, Minneapolis, Minnesota 55455, USA*

(Dated: May 13, 2021)

## S.I. CONSTRUCTION AND DESCRIPTION OF THE PHASE DIAGRAM IN FIGURE 1 OF THE MAIN TEXT

A phase diagram that describes the evolution of magnetic and electronic transport properties of $La_{1-x}Sr_xCoO_3$ is shown in Fig. 1 of the main text. The Curie-Weiss temperature ($\theta_{CW}$) was determined from fits to the susceptibility of single crystals at temperatures high enough that the susceptibility followed the Curie-Weiss law, unaffected by the spin-state transition at lower temperature. The absolute value of the Curie-Weiss temperature is plotted using green circles in Fig. 1. Note that the first 3 values plotted (*i.e.* $x \leq 0.03$) have negative Curie-Weiss temperatures, while the values at higher $x$ are all positive. Thus, there is a cross-over in the sign of the Curie-Weiss temperature, as discussed in the main text. The spin-state transition temperature ($T_{SST}$, white circles) was determined for $x = 0$ as the approximate midpoint of the low temperature rise in susceptibility with increasing temperature that occurs due to the spin-state transition. At higher $x$, La NMR data on these crystals have been used to determine $T_{SST}$, revealing that the SST still occurs in the non-ferromagnetic matrix. This is often obscured in other types of measurements [1]. The temperatures at which short-range ferromagnetic clusters are first visible via Small Angle Neutron Scattering (SANS), $T^*$, are plotted as blue circles from data reported in Ref. [2–4]. At temperatures lower than $T^*$ but higher than spin-glass freezing or ferromagnetic Curie temperatures (depending upon $x$), the ferromagnetic correlations are fluctuating on $\sim$ picosecond time-scales. The spin-glass transition temperatures ($T_G$) and the Curie temperatures ($T_C$), were determined from magnetometry (e.g., [5]) and are plotted as red circles and black circles, respectively. The dashed line at $x \approx 0.04$ marks the approximate cross-over point from the regime dominated by the polarons to the glassy short-range ferromagnetic regime as discussed in the text [3]. The solid line at $x_c \approx 0.17$ marks the composition at which a metallic percolation pathway occurs [3, 5, 6].

This separates the spin-glass ground-state (low $x$ side) from the non-uniform long-range ferromagnetic, metallic state (high $x$ side). The vertical line at $x \approx 0.22$ marks the composition where the ferromagnetic, metallic ground-state becomes uniform [3, 7]. For $x > 0.30$, where single crystal data are not available, $T_C$ and $T^*$ values from polycrystalline samples have been used [2].

## S.II. DOPING EVOLUTION OF THE LOW TEMPERATURE HEAT CAPACITY

The $x$-dependent heat capacity ($C_p$) at $T = 7$ K of single crystals of $La_{1-x}Sr_xCoO_3$ is shown in Fig. 2 of the main text. The temperature-dependent $C_p(T)$ behavior of $LaCoO_3$ and $La_{1-x}Sr_xCoO_3$ have been extensively discussed in Refs. 3, 8, and 9. Here, we discuss the evolution shown in Fig. 2 of the main text within the context of these previous measurements. In order to do so we first summarize what has previously been elucidated.

### S.A. Summary of previous measurements

#### S.1. Parent compound $x = 0$

$C_p = AT^{-2} + \gamma T + \beta T^3 + C_{SA}$ at this $x$ [8]; $C_p$ consists of a very weak electronic contribution ($\gamma T$), a lattice contribution ($\beta T^3$, where $\beta = \frac{12\pi^4}{5}\frac{n_i k_B}{\Theta_D^3}$ with ionic density $n_i$ and Debye temperature $\Theta_D$), and a Schottky Anomaly (SA). The SA is attributed to a dilute concentration of non-zero $Co^{3+}$ spin states that have been stabilized around oxygen vacancies [8]. Its $C_p$ is given by:

$$C_{SA} = \frac{n}{N_{Co}} R \left(\frac{\Delta E}{k_B T}\right)^2 \frac{\nu_0}{\nu_1} \frac{\exp(\Delta E/k_B T)}{\left[1 + \frac{\nu_0}{\nu_1} \exp(\Delta E/k_B T)\right]^2}, \quad \text{(S.1)}$$

where $n/N_{Co}$ is the fraction of involved Co sites, the degeneracies of the lower and excited energy levels are given by $\nu_0$ and $\nu_1$, and $\Delta E$ is the energy level splitting [8]. The $AT^{-2}$ term is needed to fit the lowest temperature data ($T < 1$ K) and represents the internal field arising from the $Co^{59}$ hyperfine splitting [8, 9].

* porth@iastate.edu

### S.2. Doping range $0.05 \leq x \leq 0.15$

$C_p = AT^{-2} + \gamma T + BT^2 + \beta T^3$ in this regime [9]; The SA disappears because of the stabilization of finite $T=0$ spins through Sr substitution. The electronic contribution is still nearly zero, and a new term $BT^2$ is required to fit the data. The origin of the $BT^2$ term is not entirely clear but it has been argued strongly that it arises from magnetic phase separation, i.e., short-range ferromagnetic clusters embedded in a non-ferromagnetic matrix [9]. The $AT^{-2}$ term rises due to the increasing internal field.

### S.3. Doping range $0.15 < x \leq 0.22$

$C_p = AT^{-2} + \gamma T + BT^2 + \beta T^3$ in this regime [9]; The electronic contribution increases and the $BT^2$ term diminishes to zero.

### S.4. Doping range $x > 0.22$

$C_p = AT^{-2} + \gamma T + \beta T^3$ in this regime [9]; The electronic contribution saturates and the $BT^2$ term is no longer required.

### S.B. Results from new measurements

We performed $C_p$ measurements on lower-doped crystals ($0 \leq x \leq 0.05$) in order to gain insight into the evolution from the regime described in Sec. S.1 to the regime described in Sec. S.2. Figure S1(a) shows $C_p/T$ vs $T^2$ for $x = 0, 0.005, 0.03$, and $0.05$. First, for $x = 0$ we demonstrate the refitting of LaCoO$_3$ as described above, as shown in Fig. S1(b). Values for the fitted parameters and shapes of the individual terms are listed and shown in the figure. The slight doping of $x = 0.005$ in Fig. S1(c) significantly alters the low temperature $C_p$. The shape of the temperature dependence is reminiscent of $x = 0$, but the magnitude of $C_p$ is increased. To explain the increase in $C_p$, we consider the manifold of states of an individual spin-state polaron. Podlesnyak *et al.* [10] considered that the ground-state multiplet of the seven-site polaron consists of all 7 spins aligned (six $S=1$ spins and one $S=1/2$ spin) such that the total $S=13/2$. This $S=13/2$ state is then split by a trigonal crystal field into 7 Kramer's doublets ($m=\pm 13/2$, $m=\pm 11/2$, $m=\pm 9/2$, $m=\pm 7/2$, $m=\pm 5/2$, $m=\pm 3/2$, and $m=\pm 1/2$) with $m=\pm 13/2$ as the ground-state. Assuming $H=DS_z^2$, the energy-level scheme can be deduced given an observed transition at $\Delta E = 0.75$ meV attributed to a transition from $m=\pm 13/2$ to $m=\pm 11/2$ by inelastic neutron scattering [10]. The thermal transitions within this manifold give rise to a source of excess

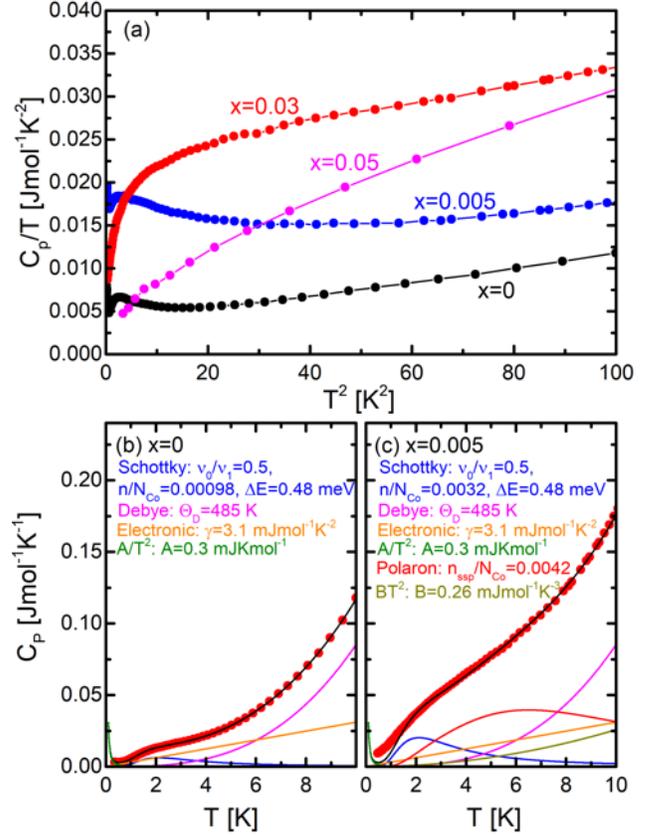

FIG. S1. (a) $C_p/T$ vs $T^2$ as measured on La$_{1-x}$Sr$_x$CoO$_3$ single crystals. (b,c) Fitted contributions to $C_p$ of LaCoO$_3$ and La$_{0.995}$Sr$_{0.005}$CoO$_3$, respectively. The listed parameters correspond to the terms described in the Supplemental text.

$C_p$, $C_{\text{SSP}}$ (SSP = spin-state polaron) given by:

$$C_{\text{SSP}} = \frac{n_{\text{SSP}}}{N_{\text{Co}}} \frac{R}{(k_B T)^2} \left( \langle E_j^2 \rangle - \langle E_j \rangle^2 \right), \quad (\text{S.2})$$

where $n_{\text{SSP}}/N_{\text{Co}}$ is the fraction of Co sites at the center of a spin-state polaron (SSP) and $\langle \cdot \rangle$ indicate a thermal average over all states.

Figure S1(b) shows $C_p$ for LaCoO$_{3-\delta}$ that is refit using $C_p = AT^{-2} + \gamma T + \beta T^3 + C_{\text{SA}}$ following Ref. [11]. Noting that this provides a good fit for $x = 0$, we employ these values as starting parameters and add the additional $C_{\text{SSP}}$ term to fit the $x = 0.005$ crystal using: $C_p = AT^{-2} + \gamma T + BT^2 + \beta T^3 + C_{\text{SA}} + C_{\text{SSP}}$, as shown in Fig. S1(c). Critically, to ensure reasonable parameters, we have constrained the Debye temperature $\Theta_D$ (and thus $\beta$), $A$, $\gamma$, $\Delta E$ and ($\nu_0/\nu_1$) to the values obtained for LaCoO$_3$. This leads to a quite reasonable fit of $C_p$ above 2 K [see Fig. S1(c)]. It also yields a reasonable value of $n_{\text{SSP}}/N_{\text{Co}}$ for $x = 0.005$ crystals ranging from 0.003 to 0.0042 (it is expected to be $x$ in the dilute limit). This lends credence to the polaron energy-level diagram [10] with the following qualifiers: (i) it fails to reproduce excess heat capacity present below 2 K, the ori-

gin of which remains unclear; (ii) it consistently yields approximately three times higher contribution of $C_{SA}$ than for $x = 0$, which may be related to some concentration of oxygen vacancies in the doped crystals. Such vacancies are very likely to form with increasing $x$, but we note that transport data on these crystals reveals monotonically decreasing resistivity with increasing $x$, even at $x$ as low as 0.005. This rules out a high concentration of oxygen vacancies, which would shift the doping compensation point from $x = 0$. Thus, we argue that the increase in the low temperature $C_p$ for $x = 0.005$ [compare $x = 0$ and 0.005 in Fig. S1(a)] is mostly due to the formation of magnetopolarons.

As $x$ is increased to $x = 0.03$ in Fig. S1(a), there is significantly more excess $C_p$ except at the very lowest temperatures. This is consistent with an increase in the number of magnetic states at low temperature, which implies that there is a higher polaron concentration. However, and importantly, the excess $C_p$ is rather structureless in temperature and can no longer be fit by that predicted for a single isolated magnetopolaron; rather, it is suggestive of polarons with highly broadened levels. Notably, further increasing $x$ to 0.05 actually leads to *reduced* $C_p$. An explanation for this is that the concentration of magnetopolarons is reduced as the hole concentration is increased; the frozen state that remains has low temperature $C_p$ dominated by the $BT^2$ term.

Now let us consider the evolution of $C_p$ at 7 K which appears in Fig. 2 of the main text. For $x = 0$, the largest contribution (beyond the Debye component) is the small electronic term. However, for $x = 0.005$, there is a large additional contribution from the polarons; this explains the increase up to $x = 0.03$. There is then a drop from $x = 0.03$ to $x = 0.05$ as the term from the polarons drops out, and the $BT^2$ term becomes dominant. At percolation ($x \approx 0.17$), the electronic component becomes large and an increase occurs. This is mirrored in the ferromagnetic volume fraction (also shown in Fig. 2 of the main paper), which was estimated by extrapolating the high field magnetizations vs. field behavior back to zero field, and normalizing to the value deep in the ferromagnetic phase (where the ferromagnetism is uniform).

## S.III. MONTE-CARLO RESULTS FOR THE FERROMAGNETIC BINDER CUMULANT

Figure S2 shows the (disorder averaged) ferromagnetic Binder cumulant [12, 13]

$$U_F = \frac{3}{2}\Big(1 - \frac{1}{3}\frac{[\langle m^4 \rangle]_{\text{dis}}}{[\langle m^2 \rangle]^2_{\text{dis}}}\Big) \quad (S.3)$$

as a function of temperature $T$ for different system sizes $L$. Different panels show results for different $Co^{4+}$ doping levels $x = 0.5$ (a), $x = 0.3$ (b), $x = 0.25$ (c), and $x = 0.2$ (d). Here, $m = \frac{1}{N}\sum_{i=1}^{N} S_i$ is the F magnetization and $N = L^3$. The crossing of $U_F$ for different system sizes is used to extract the F transition temperatures used in Fig. 4(a) of the main text. We find that as $x$ is reduced, $T_F$ decreases monotonically. Notably, for $x < 0.25$, there is no single crossing point of $U_F$ for the three system sizes used. Therefore, and because we find peaks in both F and AF susceptibilities that only moderately grow with system size, we identify the region $x < 0.25$ as a magnetic glassy phase. The Monte-Carlo simulation parameters are given in the main text.

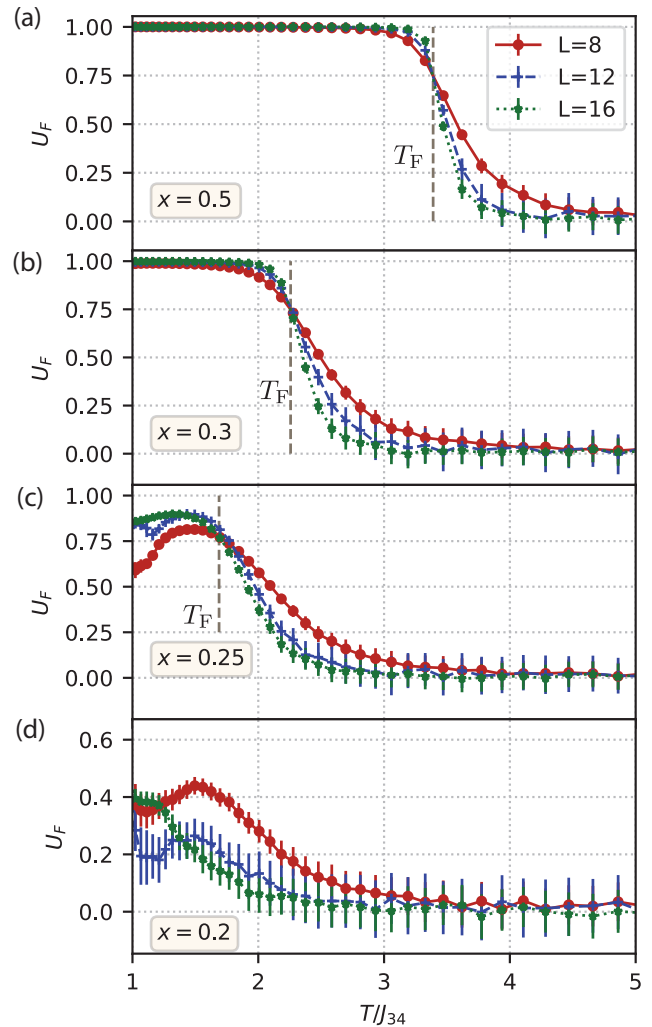

FIG. S2. Ferromagnetic Binder cumulant $U_F$ as a function of $T/|J_{34}|$ for different system sizes $L = 8, 12, 16$. Different panels show different $Co^{4+}$ doping levels $x = 0.5, 0.3, 0.25, 0.2$ (a-d). The crossing of $U_F$ for different system sizes $L$ is used to extract the F transition temperatures $T_F(x)$ that are shown in Fig. 4(a) of the main text. Crossing of Binder cumulant is absent for $x < 0.25$, which we identify as a magnetic glassy region.



\end{document}